\documentclass[a4paper,11pt]{elsarticle}

\begin{document}

\begin{frontmatter}

\title{A Double Sided LGAD-Based Detector Providing Timing, Position, and Track Angle Information}


\author{Ronald Lipton \\
Fermlab Report: FERMILAB-FN-1102-E}{\corref{mycorrespondingauthor}}
\address{Fermilab, PO Box 500, Batavia, Il 60510}
\cortext[mycorrespondingauthor]{Corresponding author}
\ead{lipton@fnal.gov}

\begin{abstract}
We describe a detector concept which combines small pixels enabled by 3D sensor-electronics integration with gain 
produced by a Low Gain Avalanche Diode (LGAD) layer. The detector is double-sided, with electrons collected by the 
cathode, which provides timing information, and an anode with small pixels to provide position and angle information. 
The cathode can be coarse grained, providing timing with fewer fast amplifiers to limit power consumption. The anode 
layer benefits from the gain of the LGAD, with larger signals that also limit the power needed.  Position can be 
reconstructed by measuring the pattern of total charge deposited on the anode. Angle and depth of charge deposit 
can be measured by the shapes of the anode pulses. We describe a possible assembly technology based on  
3D integration.
\end{abstract}

\begin{keyword}
3D Integration \sep Pixel Detector \sep Induced Currents \sep Silicon Tracking
\end{keyword}

\end{frontmatter}

\section{Introduction}
Low Gain Avalanche Diodes (LGADs) have become primary candidates for fast timing tracking applications in particle physics \cite{SADROZINSKI201618}
\cite{0034-4885-81-2-026101}\cite{SADROZINSKI20147}\cite{SADROZINSKI2013226}\cite{1748-0221-11-12-C12016}.
These devices achieve moderate internal gain in silicon-based detectors by adjusting doping and bias potential to achieve high internal fields.  
LGADs have been chosen for timing layers in both the ATLAS and CMS HL-LHC upgrades \cite{Collaboration:2296612} \cite{Collaboration:2719855}.
The current generation of devices suffer from low fill factor for small pixels due to large junction termination structures. These 
problems are being solved by incorporation of an AC coupling layer or trench isolation of pixels \cite{CARTIGLIA2015141}\cite{Tornago:2020otn}
\cite{Apresyan:2020ipp}\cite{Giacomini:2019kqz}\cite{Pellegrini:2015hug}.  The use of these structures 
provides the possibility of incorporating small pixels ($25 \mu $ or less) into an LGAD structure. This, in turn, allows us to consider 
the possibility of a single "5D" detector with excellent timing, position, and track angle resolution.

In a previous note we discussed the possibility of a 5D detector utilizing induced currents in small pixels \cite{Lipton:2019drv}. This device can provide both 
time and angle resolution. However such a device requires many pixels with fast timing capability and complex readout to disentangle 
the complex patterns of induced current flow.  Here we present an alternative device which uses a double sided LGAD-based detector 
to provide separate timing, position and angle information.

\section{Concept}
The basic concept of the 5DLGAD consists of  a double-sided silicon detector with a gain layer on the electron-collecting side and an array of 
small electrodes on the hole-collecting (anode) side. We assume that the detector is thick compared to the anode pitch.
The electron-collecting cathode will observe a fast rise-time signal due to the avalanche in the nearby  gain 
layer. The electrodes on the hole-collecting side will observe a current initially due to the direct track ionization and then due to the holes generated 
in the gain layer.  Because the anode is more finely segmented than the cathode the weighting field plays a significant role and the 
induced current in the segmented anodes will be dominated by local charge motion. The resulting current shapes on the anodes 
reflect the pattern and depth of charge deposition and therefore the angles of incident tracks.
For example, an inclined track this can resolve into a double pulse with a separation in time proportional to the depth of the charge 
deposit in the silicon.
This should allow us to extract the angle of the inclined track by the pattern of currents in the pixel
array.

 \begin{figure}[ht]
\centering
\includegraphics[width=100 mm]{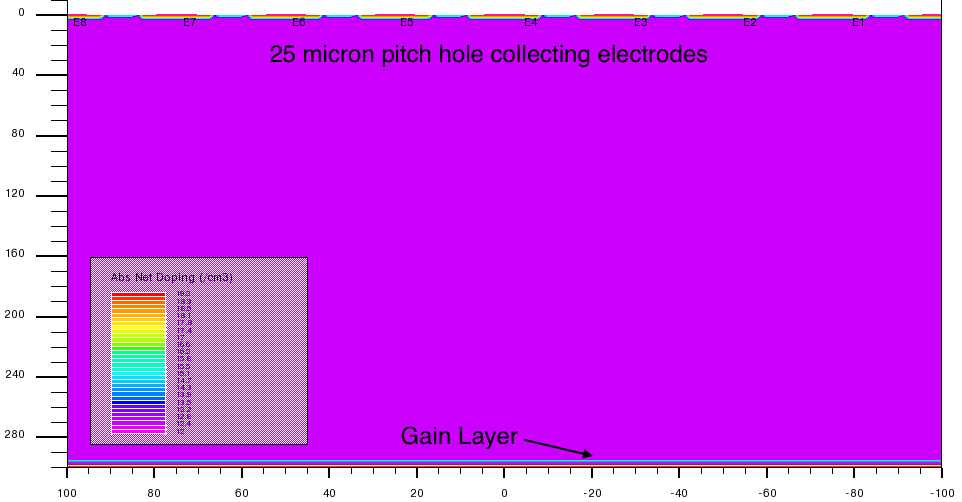}
\caption{Simulated detector structure. The detector is 300 microns thick with 25 micron pitch hole-collecting ohmic electrodes on the top and 
a single junction electrode on the bottom collecting electrons. The gain region is 3-5 microns above the junction implant}
\label{str}
\end{figure}

\section{Simulation}
\begin{figure}[ht]
\centering
\includegraphics[width=60 mm, trim={0 0 0 2.5cm},clip]{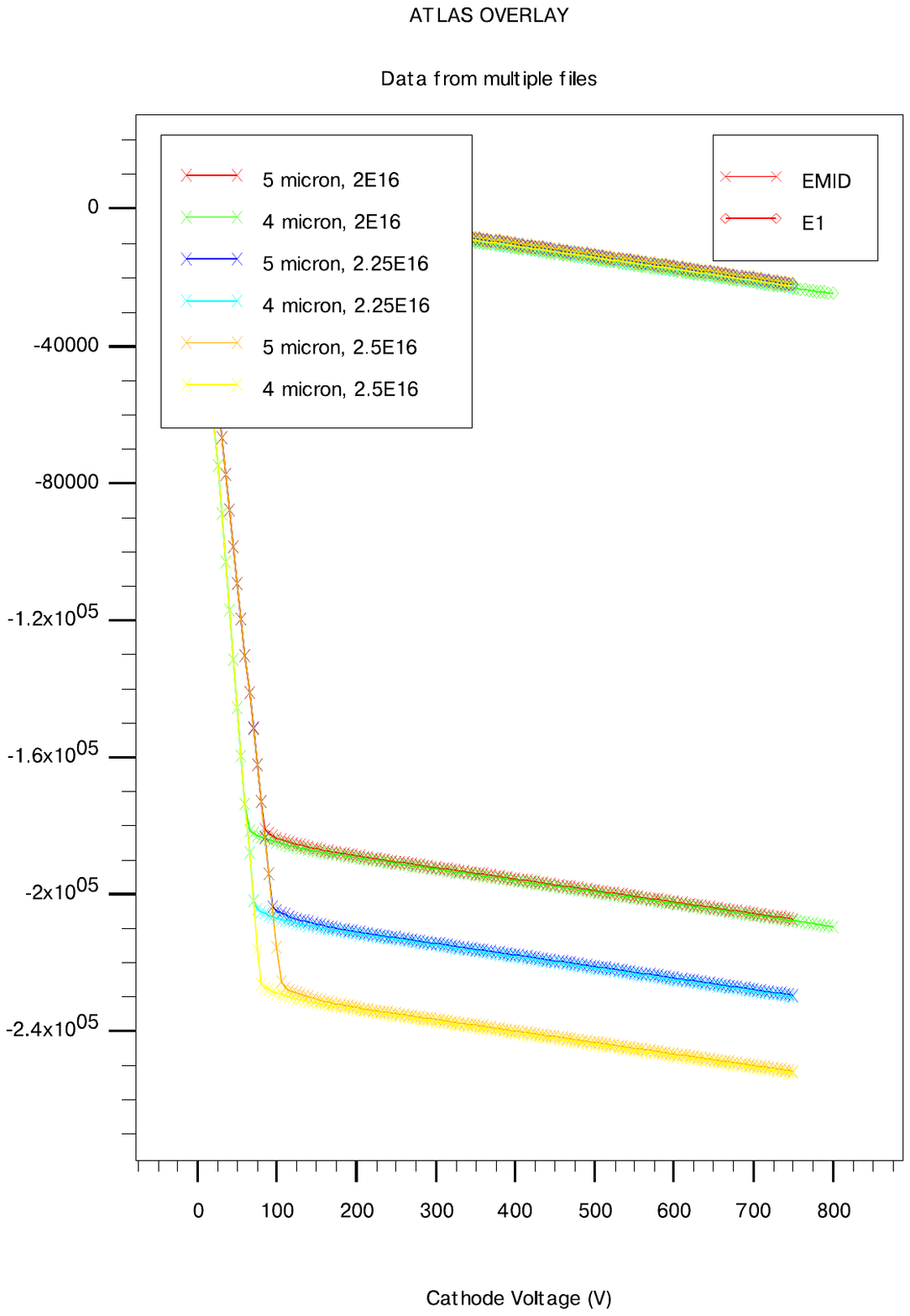}
\includegraphics[width=60 mm, trim={0 0 0 2.5cm},clip]{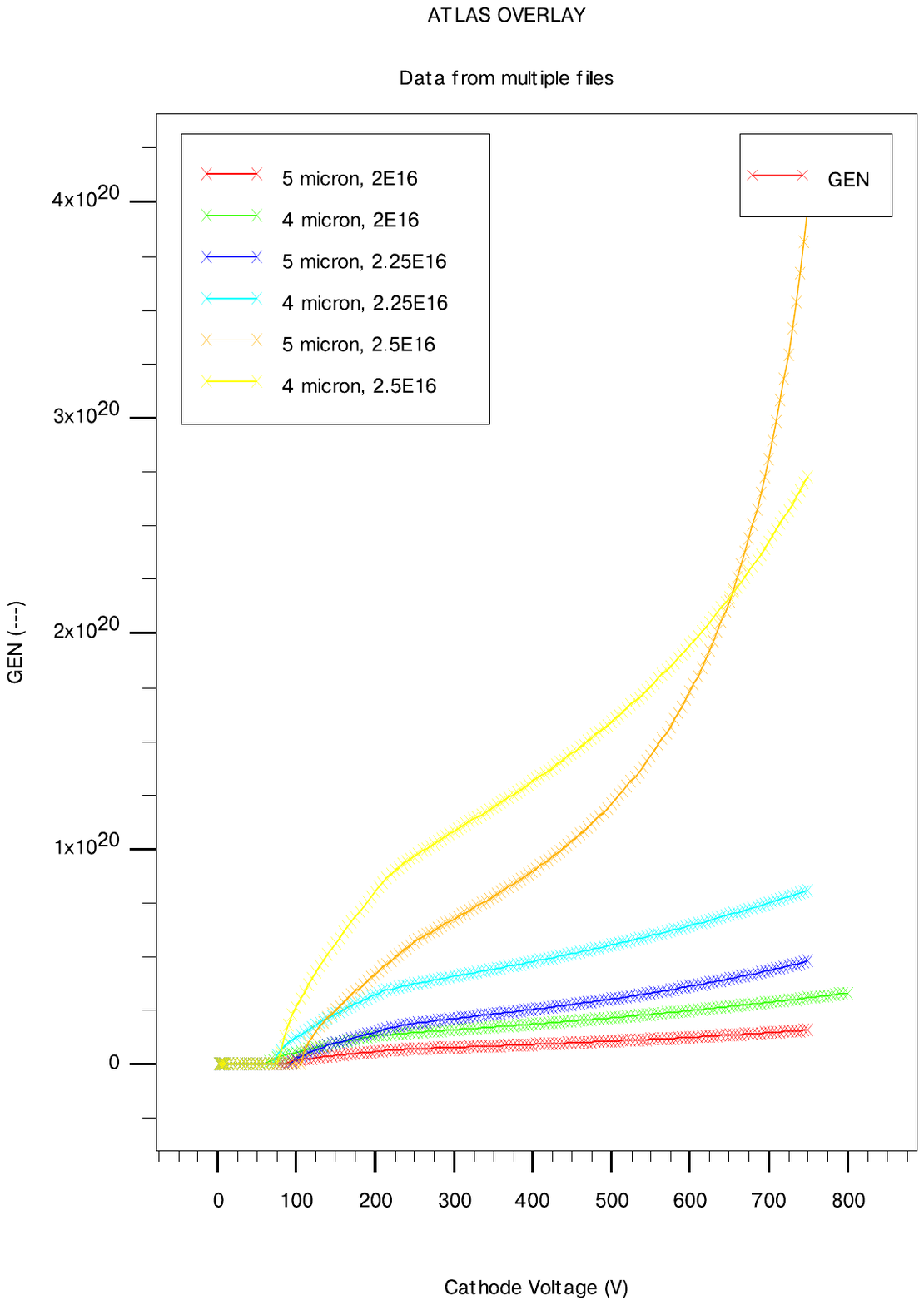}
\caption{Left - Drift (E1) and gain region (EMID) fields as a function of bias voltage for various gain layer peak implants. Right - Gain layer impact generation 
rate as a function of bias voltage for various gain layer conditions}
\label{gainl}
\end{figure}

We used Silvaco TCAD to simulate the candidate detector structure.  The structure shown in figure ~\ref{str} is a p-bulk detector with 25 micron 
pitch p-on-p pixels on the top and a single junction electrode on the bottom. Gain is achieved with a buried p-type gain layer approximately 2.5 microns 
above the cathode n-type implant.  This is assumed to be a buried gain layer achieved either by wafer bonding or by a graded epitaxial layer.
To achieve a workable candidate design we need to define the buried layer doping density such that the layer can be fully depleted at the the 
operating voltage with a gain field giving a reasonable gain (~10) while also giving a large enough field in the drift region to collect the holes with 
acceptable speed and diffusion. Figure \ref{gainl} shows a study showing the tradeoff between gain layer doping and drift and gain fields. The drift field
is zero until the gain layer is fully depleted, during this time the gain field increases rapidly due to the small gap between the gain layer and the cathode. 
 Once the the gain layer is depleted the plates of the virtual capacitor separate, and both the drift field and the gain field increase at a 
 rate that is now determined by the full detector thickness, in this case 300 microns. 
 
 In our simulation we force the cathode current until the leakage current reaches the desired value, indicating avalanche multiplication of the 
 leakage current.  This is about 10 na  for most of our simulations. 
 We can monitor the effective gain by extracting the impact generation rate in the gain region from the 
 simulation. Gain layer depth and doping can be tuned to achieve the desired gain characteristics. The 
 gain itself can be determined by calculating the ionization integral for the desired configuration.
 
For initial studies, we chose a configuration with relatively low gain with a gain layer depth of 4 microns, doping of $2.25 \times 10^{16}$ operated at 
a bias determined by the 10 na leakage current.  This corresponds  to a gain of 3-4.  The field in the drift region is about $21,000$ V/cm.  
This choice in not optimized and will be revisited for detailed design studies, but it provides a demonstration of the effect.

\section{Sensor Current Pulse Characteristics}
\begin{figure}[ht]
\centering
\includegraphics[width=100 mm, height=70mm, trim={0 0 0 2.5cm}, clip]{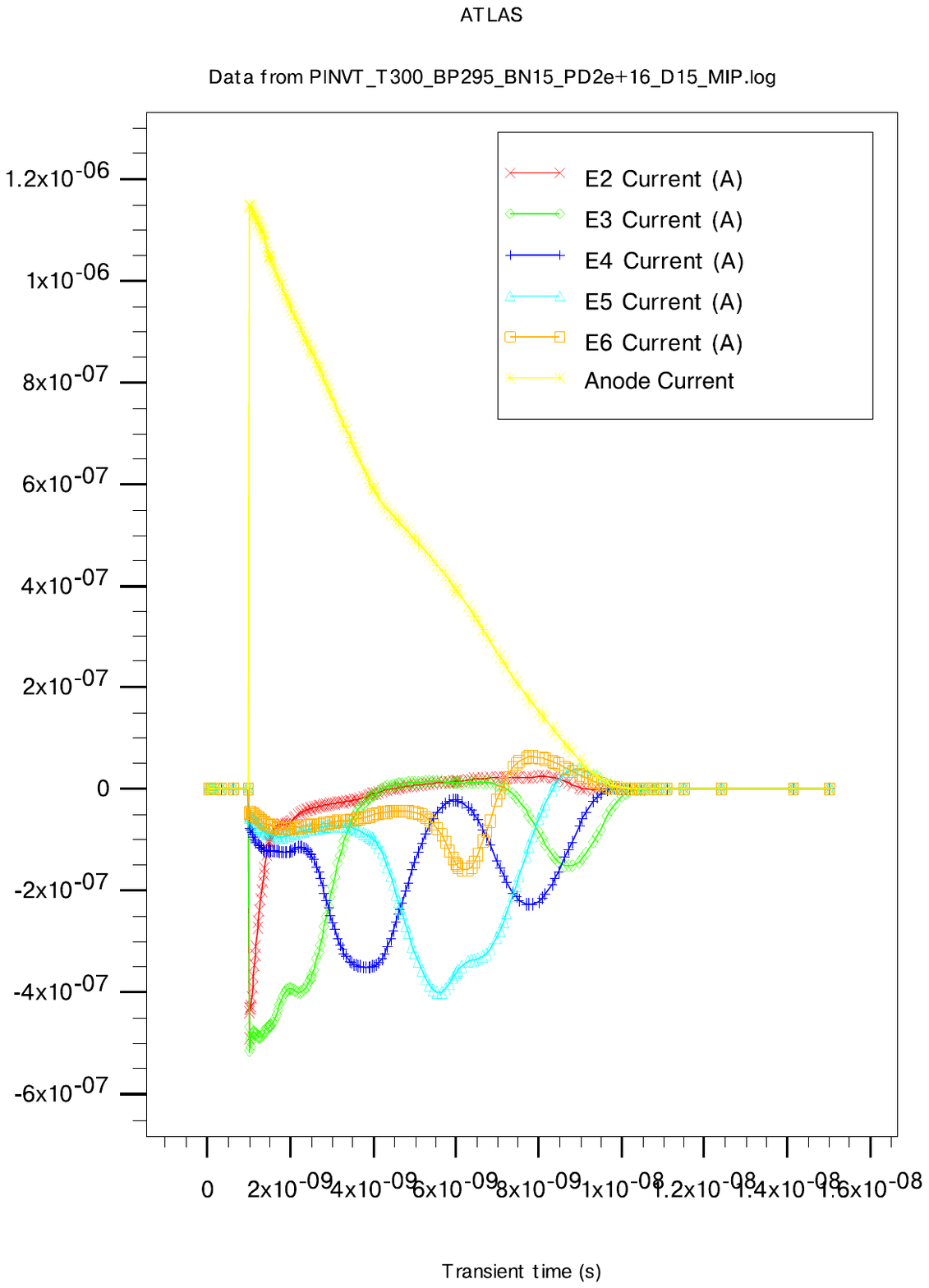}
\caption{Pulse shapes resulting from a 15 degree track for the cathode and five of the anodes. Charge deposited near the electrode 3 anode (green) shows 
the initial induced charge from direct collection of holes and a delay corresponding to electron drift to the gain layer followed by hole drift back to the anode.}
\label{15deg}
\end{figure}

We examine the transient response of the detector by injecting electron-hole pairs corresponding to a minimum ionizing track at various incident angles. 
The track is assumed to cross the midplane of the detector at x=0. Figure \ref{15deg} shows the currents in the middle 5 electrodes for a track incident 
at a 15 degree angle at t=1 ns. The electron-collecting cathode pulse has a very fast rise time and large signal. Electrodes 3 and four show a double 
pulse structure, the first due to collection of the direct hole ionization and the second due to the collection of holes amplified in the gain layer. 
The position of the first peak is determined by the hole drift to the top electrode.  The position of the second peak corresponds to the electron drift 
to the gain layer plus the hole drift to from the gain layer to the top electrode.
Charge deposited below electrodes 5 and 6 is sufficiently deep not to induce a significant current due to primary ionization, but show clear signals 
from holes generated in the gain layer. These current relations correspond to depth of charge deposit in the silicon and can be unfolded to 
reconstruct the incident angle of the track.

\begin{figure}[ht]
\centering
\includegraphics[width=130mm]{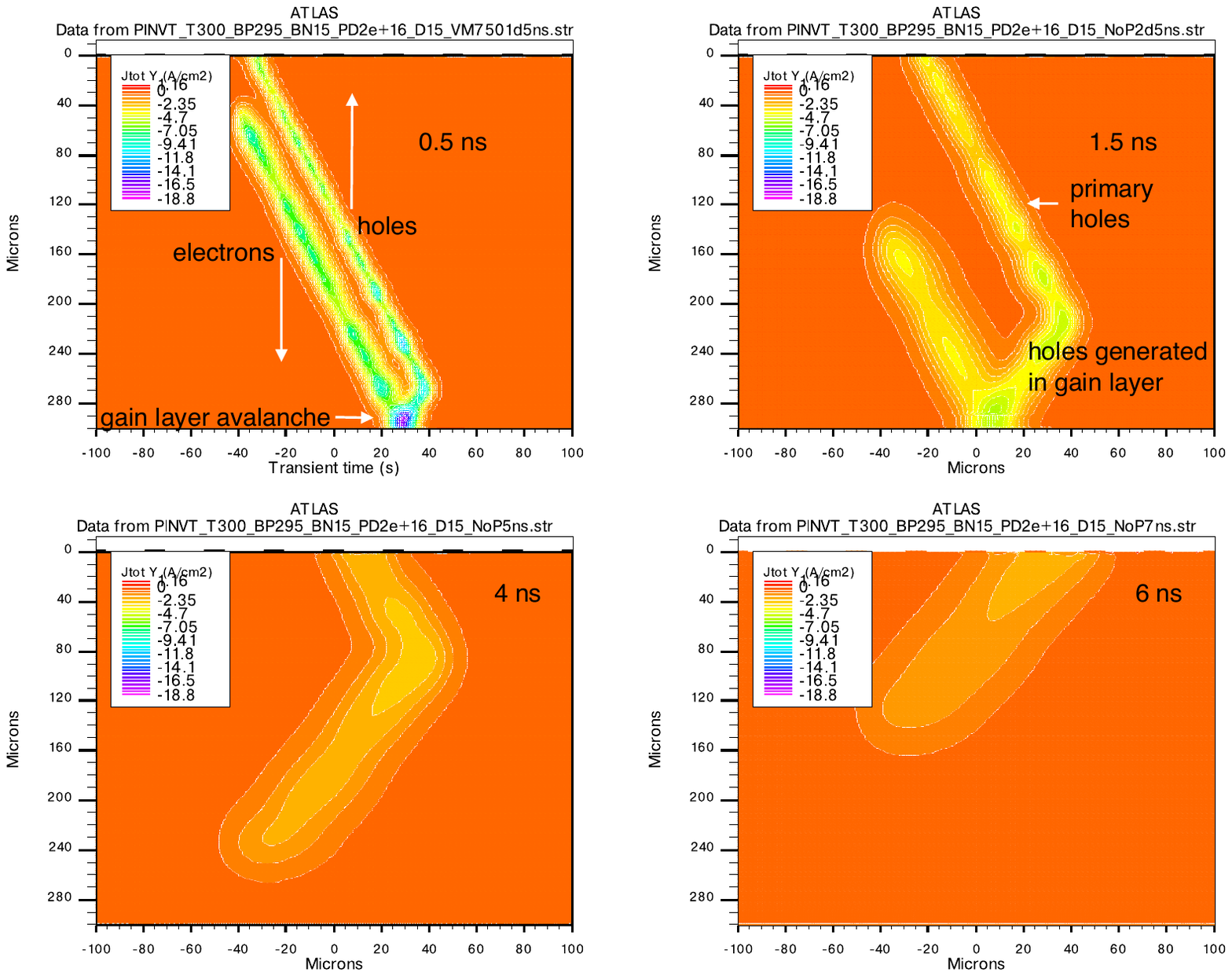}
\caption{Snapshots of summed vertical electron and hole currents at various times after the initial charge deposition. }
\label{carriers}
\end{figure}

Figure \ref{carriers} shows a set of snapshots of the currents in the silicon detector.  At 0.5 ns electrons below 275 microns have been collected and we see 
holes from the gain layer beginning to drift up. At 1.5 ns the electrons and holes are well-separated with the electron cloud extending from 150-300 microns with 
noticeable diffusion. We both the primary holes drifting up as well as the holes from the gain layer forming a "tail". At 4 ns all of the electrons have been 
collected and the current dogleg is composed of primary holes from 120-0 microns and gain holes from 120-300 microns.  Finally at 6 ns all of the primary 
holes have been collected and we see only the tail of the gain layer-generated holes.

For a charge deposit in the silicon bulk the pulse timing can be qualitatively described, ignoring details of the the gain region at the cathode, the actual 
shape of the amplified pulse at the gain layer, and charge diffusion effects as initial and final pulses:
\begin{eqnarray}
T_{inital} = \int_{Ydep}^{Yanode} v_{dr,h}(\overrightarrow{E}) dy \\
T_{final} =  \int_{Ycathode}^{dep} v_{dr,e}(\overrightarrow{E} ) dy + \int_{Ycathode}^{Yanode} v_{dr,h}(\overrightarrow{E} ) dy
\end{eqnarray}
Where $ v_{dr,h}(\overrightarrow{E})$ is the hole drift velocity and $ v_{drift,e}(\overrightarrow{E})$ is the electron drift velocity, $Y_{dep}, Y_{anode} $ and 
$Y_{cathode}$, are the depths of the charge deposit, anodes (nominally at Y=0), and cathode. The drift field ($\overrightarrow{E}$), in the baseline 
simulation ranges from 18,000 V/cm near the anodes to 26,000 V/cm near the cathodes before the gain region. 

\section{Time Resolution}
The time resolution should be similar to a standard LGAD, but because the substrate is thick, the device will be more sensitive to fluctuations in 
ionization along the particle path then the more standard 50 micron thick devices.  Studies of time resolution vs LGAD thickness show 
fluctuation-limited resolution of about 75 ps for a 300 micron device\cite{SADROZINSKI201618}.  This may be improved 
by optimization of the anode design and segmentation of the cathode to provide lower capacitance, but this device is unlikely to achieve the 20-30 ps per pixel resolution expected for a $50 \mu$m thick LGAD.

\section{Angular Reconstruction}
\begin{figure}[ht]
\centering
\includegraphics[angle=-90, width=120mm]{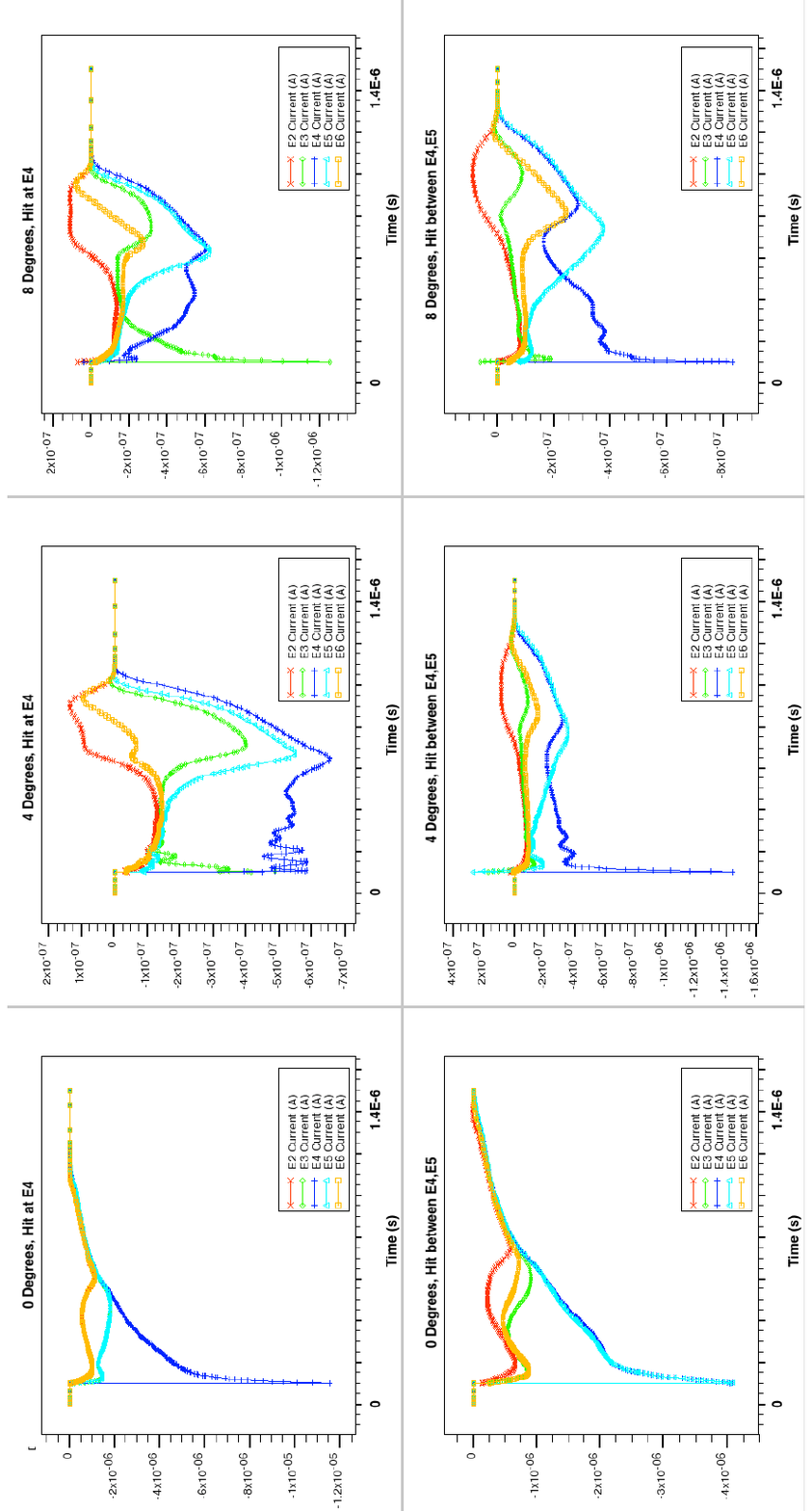}
\caption{Pulse shapes resulting from tracks at 0, 4, and 8 degrees for track impact midpoints (z=150 microns) at 0 and 12.5 microns - 
between electrodes 4 and 5. The rapid oscillation in some shapes are due to the size of the TCAD mesh.  Current values are 
not normalized.}
\label{IP-comp}
\end{figure}

 The DSLGAD has intrinsic angular discrimination due to the pulse shape dependence on the depth of charge deposition. 
 Individually the pulse shapes can be rather complex, due to the time dependence of electron and hole motion. These shapes 
 also depend on the impact point of the track. This is demonstrated in figure \ref{IP-comp}  which shows pulse shapes for 
 tracks at 0, 4, and 8 degrees impacting the detector vertical midplane at 0 (the center of electrode 4) and 12.5 microns (between 
 electrodes 4 and 5).  There are systematic variations with angle and position, but the overall pattern is complex and full reconstruction 
 of individual track angle and position may require fairly sophisticated processing and may be difficult to achieve on-chip.
 
\begin{figure}[ht]
\centering
\includegraphics[angle=-90, width=100mm]{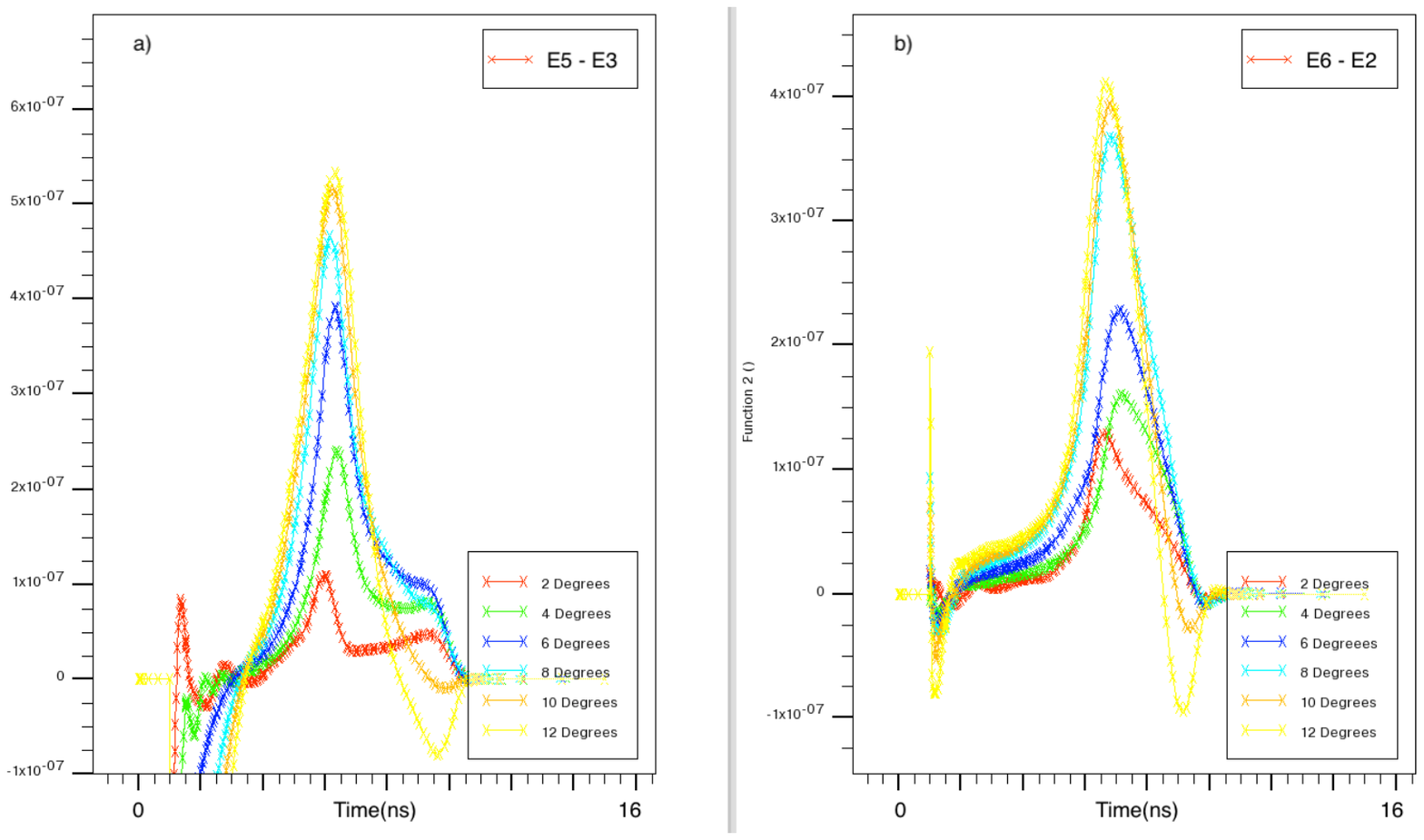}
\caption{Subtracted pulse shapes for neighbor (a, E5-E3)) and next-neighbor electrodes (b, E6-E2)) in the case 
where the track impacts the midplane of the detector at the center of electrode 4 (E4). }
\label{angles}
\end{figure}

In some circumstances the pattern is simpler.   For example, if we 
 take the difference between pads centered around the charge deposit we see a simpler shape that has a peak around the hole transit time
 of about 6.8 ns for our sample configuration and monotonic dependence on track angle. Figure \ref{angles} shows neighbor (electrode 3  - electrode 5) and
next-neighbor ( electrode 2 - electrode 6) 
 subtracted pulse shape distributions for a track centered on electrode 4.  This technique has the advantage that it can be simply 
implemented by using differential amplifiers on neighbor pads.

This is a simplified example with 
no lorentz force or delta rays. On-line angle resolution 
is likely to be determined by variations in impact point with respect to the electrode center. There will also be errors due to fluctuations in track ionization 
and gain layer amplification. The range of angles with optimum resolution can be adjusted by changing the width/pitch of the electrodes.  The case of a fixed 
impact point  
there is good resolution (probably around a degree) below 8 degrees with coarser resolution beyond 10 degrees.  Offline the region of good angular 
resolution can be extended by analyzing the full pattern of electrode pulse heights. 

Our device simulation is two dimensional.  This is appropriate for an application such as the CMS two-layer trigger module where we are interested in one 
dimensional track bend information. This application also allows for more available space for analog processing and digital trigger and readout electronics. 
For applications where we are interested in the 2D angle, such as background rejection for a muon collider, the device can be pixelated.  This is possible 
because of the large hole signal provided by LGAD amplification. The large LGAD hole signal also means that less power is needed for the front-end amplifier. 
The overall design of such a device will depend very much on the final application.

\section{Fabrication Options}
\begin{figure}[ht]
\centering
\includegraphics[width=70mm]{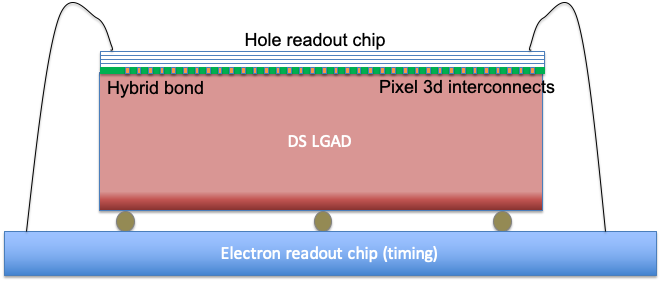}
\includegraphics[width=40mm]{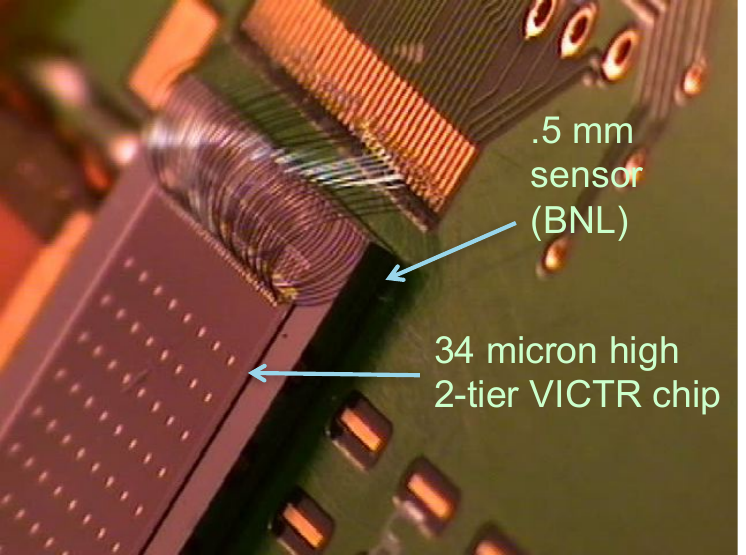}
\caption{Left - Schematic of a possible 5D LGAD detector assembly. The top readout ASIC is hybrid bonded to the LGAD sensor with fine pitch contacts to the 
top readout electronics. The bottom, bump bonded, contacts provide bias and connection to the timing amplifiers and digitizers. Right - Photograph of the 
VICTR 3D chip bonded to a 0.5 mm thick sensor. The VICTR is thinned to 35 microns and bump pads deposited on top.  These were later connected to a 
second sensor to provide 2 layer track correlations. }
\label{assy}
\end{figure}

The DSLGAD depends on the ability to read out a double sided detector with fine pitch pixels.  3D integration of the detector and electronics allows for (almost)
straightforward construction of such a device.  Figure \ref{assy} shows a possible geometry. It utilizes a fine pitch hybrid (3D) bonded 
top ASIC for hole readout and more conventional bump bonds to the bottom chip which reads out electrons and provides timing. We demonstrated a 
similar geometry in our VICTR 3D project in 2012 \cite{lipton2017}\cite{7167712}
 which included a combination of 3D hybrid bonds to a base sensor and bump bonds to a 
top sensor tier. 
Hybrid bonding allows thinning of the top ASIC and contact pads can be deposited directly on the top surface. This was done for the VICTR chip. The cathode (bottom) 
contact is much coarser pitch than the envisioned 25 micron pitch top readout.  Bias is applied to the bottom electrode, which must be AC coupled.
The is the most significant technical challenge in the assembly.  An oxide layer, similar to that used in AC LGADs, might provide sufficient dielectric 
strength. An alternative construction  would capacitively couple the sensor and readout chip using a thin glue layer as the dielectric.  This 
technique has been demonstrated for the CLICPIX and FEI4 chips\cite{TEHRANI:2016}.

\section{Conclusions} 
We have presented a concept for a doubled sided LGAD-based detector where the cathode can provide coarse pitch timing information and the anodes can be 
segmented to provide position and angle information. The separated functions in such a device can be tailored for the the required application by engineering 
the thickness, electrode configuration, and internal fields. If timing is not necessary, the DSLGAD can still provide precise position resolution and 
approximate track angle information. A DSLGAD may be especially useful for tracking in a muon collider where there is a very high background of 
off-angle tracks.
\section{Acknowledgments}
This manuscript has been authored by Fermi Research Alliance, LLC under Contract No. DE-AC02-07CH11359 with the U.S. Department of Energy, Office of Science, Office of High Energy Physics.

\section{References}

\bibliography{DSLGAD}

@techreport{Collaboration:2296612,
      author        = "Collaboration, CMS",
      title         = "{TECHNICAL PROPOSAL FOR A MIP TIMING DETECTOR IN THE CMS
                       EXPERIMENT PHASE 2 UPGRADE}",
      institution   = "CERN",
      address       = "Geneva",
      number        = "CERN-LHCC-2017-027. LHCC-P-009",
      month         = "Dec",
      year          = "2017",
      reportNumber  = "CERN-LHCC-2017-027",
      url           = "http://cds.cern.ch/record/2296612",
      note          = "This document describes a MIP timing detector for the
                       Phase-2 upgrade of the CMS experiment, in view of HL-LHC
                       running",
}

@techreport{Collaboration:2719855,
      author        = "Collaboration, ATLAS",
      title         = "{Technical Design Report:  A High-Granularity Timing
                       Detector for the ATLAS Phase-II Upgrade}",
      institution   = "CERN",
      address       = "Geneva",
      number        = "CERN-LHCC-2020-007. ATLAS-TDR-031",
      month         = "Jun",
      year          = "2020",
      reportNumber  = "CERN-LHCC-2020-007",
      url           = "http://cds.cern.ch/record/2719855",
}

@article{CARTIGLIA2015141,
	Author = {N. Cartiglia and others},
	Journal = {Nuclear Instruments and Methods in Physics Research A},
	Pages = {141 - 148},
	Title = {Design optimization of ultra-fast silicon detectors},
	Volume = {796},
	Year = {2015}}

@article{Lipton:2019drv,
    author = "Lipton, Ronald and Theiman, Jason",
    title = "{Fast Timing with Induced Current Detectors}",
    reportNumber = "FERMILAB-CONF-18-628-CMS-E-PPD",
    doi = "10.1016/j.nima.2019.162423",
    journal = "Nucl. Instrum. Meth. A",
    volume = "945",
    pages = "162423",
    year = "2019"
}

@article{0034-4885-81-2-026101,
	Author = {Hartmut F-W Sadrozinski and Abraham Seiden and Nicol Cartiglia},
	Journal = {Reports on Progress in Physics},
	Number = {2},
	Pages = {026101},
	Title = {4D tracking with ultra-fast silicon detectors},
	Volume = {81},
	Year = {2018}}

@article{SADROZINSKI201618,
	Author = {H.F.-W. Sadrozinski and others},
	Journal = {Nuclear Instruments and Methods in Physics Research A},
	Pages = {18 - 23},
	Title = {Ultra-fast silicon detectors (UFSD)},
	Volume = {831},
	Year = {2016}}

@article{SADROZINSKI20147,
	Author = {H.F.-W. Sadrozinski and others},
	Journal = {Nuclear Instruments and Methods in Physics Research A},
	Pages = {7 - 11},
	Title = {Sensors for ultra-fast silicon detectors},
	Volume = {765},
	Year = {2014}}

@article{Tornago:2020otn,
    author = "Tornago, M. and others",
    title = "{Resistive AC-Coupled Silicon Detectors principles of operation and first results from a combined laser-beam test analysis}",
    eprint = "2007.09528",
    archivePrefix = "arXiv",
    primaryClass = "physics.ins-det",
    month = "7",
    year = "2020"
}

@article{7167712,
	Author = {G. W. Deptuch and others},
	Journal = {IEEE Transactions on Electron Devices},
	Month = {Jan},
	Number = {1},
	Pages = {205-214},
	Title = {Fully 3-D Integrated Pixel Detectors for X-Rays},
	Volume = {63},
	Year = {2016}}

@article{SADROZINSKI2013226,
	Author = {H. F.-W. Sadrozinski and others},
	Journal = {Nuclear Instruments and Methods in Physics Research A},
	Pages = {226 - 231},
	Title = {Ultra-fast silicon detectors},
	Volume = {730},
	Year = {2013}}

@article{1748-0221-11-12-C12016,
	Author = {N. Cartiglia and others},
	Journal = {Journal of Instrumentation},
	Number = {12},
	Pages = {C12016},
	Title = {The 4D pixel challenge},
	Volume = {11},
	Year = {2016}}

@article{lipton2017,
	Author = {R. Lipton},
	Journal = {Proceedings of Science},
	Month = {November},
	Note = {(2018)},
	Title = {3D IC integration},
	Volume = {309},
	Year = {2018}}

@article{Apresyan:2020ipp,
    author = "Apresyan, Artur and Chen, Wei and D'Amen, Gabriele and Di Petrillo, Karri Folan and Giacomini, Gabriele and Heller, Ryan and Lee, Hakseong and Los, Sergey and Moon, Chang-Seong and Tricoli, Alessandro",
    title = "{Measurements of an AC-LGAD strip sensor with a 120 GeV proton beam}",
    eprint = "2006.01999",
    archivePrefix = "arXiv",
    primaryClass = "physics.ins-det",
    reportNumber = "FERMILAB-PUB-20-293-CMS",
    month = "6",
    year = "2020"
}

@article{Giacomini:2019kqz,
      author         = "Giacomini, Gabriele and Chen, Wei and D'Amen, Gabriele
                        and Tricoli, Alessandro",
      title          = "{Fabrication and performance of AC-coupled LGADs}",
      year           = "2019",
      eprint         = "1906.11542",
      archivePrefix  = "arXiv",
      primaryClass   = "physics.ins-det",
      SLACcitation   = "%%CITATION = ARXIV:1906.11542;%%"
}

@article{Pellegrini:2015hug,
      author         = "Pellegrini, G. and others",
      title          = "{Recent Technological Developments on LGAD and iLGAD
                        Detectors for Tracking and Timing Applications}",
      booktitle      = "{Proceedings, 10th International 'Hiroshima' Symposium on
                        the Development and Application of Semiconductor Tracking
                        Detectors (HSTD-10): Xian, China, September 25-29, 2015}",
      journal        = "Nucl. Instrum. Meth.",
      volume         = "A831",
      year           = "2016",
      pages          = "24-28",
      doi            = "10.1016/j.nima.2016.05.066",
      eprint         = "1511.07175",
      archivePrefix  = "arXiv",
      primaryClass   = "physics.ins-det",
      SLACcitation   = "%%CITATION = ARXIV:1511.07175;%%"
}

@article{TEHRANI:2016,
    author = "Alipour Tehrani, Niloufar",
    editor = "Bisogni, Maria Giuseppina and Grassi, Marco and Incagli, Marco and Paoletti, Riccardo and Signorelli, Giovanni",
    title = "{CLIC vertex detector R{\&}D}",
    doi = "10.1016/j.nima.2015.08.072",
    journal = "Nucl. Instrum. Meth. A",
    volume = "824",
    pages = "470--472",
    year = "2016"
}

\end{document}